\documentclass[useAMS,usenatbib]{mn2e}
\usepackage{epsfig}

\def\msun{{\rm {M}}_{\odot}}
\newcommand{\etal}{{et al.}~}
\newcommand{\eg}{{e.g.~}}
\newcommand{\ie}{{i.e.~}}
\newcommand{\apj}{{ApJ}}

\def \ltsima{$\; \buildrel < \over \sim \;$}
\def \simlt{\lower.5ex\hbox{\ltsima}}            
\def \gtsima{$\; \buildrel > \over \sim \;$}
\def \gtsima{\mbox{$\; \buildrel > \over \sim \;$}}
\def \simgt{\lower.5ex\hbox{\gtsima}}            

\title[Subhaloes]{Dark Matter Subhaloes in Numerical Simulations}
\author[Reed \etal] {Darren Reed,$^{1}$\thanks{Email:
d.s.reed@durham.ac.uk}
Fabio Governato,$^{2,3}$
Thomas Quinn,$^2$
\newauthor
Jeffrey Gardner,$^4$
Joachim Stadel,$^5$
and George Lake$^6$\\
$^1$Institute for Computational Cosmology, Dept. of Physics, University of
Durham, South Road, Durham DH1 3LE, UK\\
$^2$Astronomy Department, Box 351580, University of Washington, Seattle,
WA 98195 USA\\
$^3$INAF, Osservatorio Astronomico di Brera, via Brera 28, I-20131 Milano,
Italy\\
$^4$Pittsburgh Supercomputing Center, 4400 Fifth Avenue, Pittsburgh, PA 
15213, USA\\
$^5$Institute for Theoretical Physics, University of Zurich,
Winterthurerstrasse 190, 8057, Switzerland\\
$^6$Department of Physics, Washington State University, PO Box 642814, 
Pullman, WA 99164 USA}

\pagerange{\pageref{firstpage}--\pageref{lastpage}}
\pubyear{2004}

\begin{document}

\maketitle

\label{firstpage}

\begin{abstract}

We use cosmological $\Lambda$CDM numerical simulations to model the
evolution of the substructure population in sixteen dark matter
haloes with resolutions of up to seven million particles within the
virial radius.  The combined substructure circular velocity distribution 
function
(VDF) for hosts of 10$^{11}$ to 10$^{14}$ $\msun$ at redshifts from zero 
to two or higher
has a self-similar shape, is independent of host halo mass and 
redshift, and follows the
relation ${\rm dn/dv=(1/8)(v_{cmax}/v_{cmax,host})^{-4}}$.
Halo to halo variance in the VDF is a factor of roughly two to four. 
At high redshifts, we find preliminary evidence for fewer large 
substructure haloes (subhaloes).
Specific angular momenta are significantly lower for subhaloes nearer 
the host halo centre where tidal stripping is more effective.  
The radial distribution of subhaloes is marginally
consistent with the mass profile for r$\simgt$0.3r${\rm _{vir}}$,
where the possibility of
artificial numerical disruption of subhaloes can be most reliably excluded 
by our convergence study, although a subhalo distribution that is 
shallower than the mass profile is favoured.
Subhalo masses but not circular velocities decrease toward the host centre.
Subhalo velocity dispersions hint at a positive velocity bias at small 
radii.
There is a weak bias toward more circular orbits at lower redshift,
especially at small radii.
We additionally model a cluster
in several power law cosmologies of $P \propto k^{n}$, and demonstrate
that a steeper spectral index, $n$, results in significantly less
substructure.

\end{abstract}


\section{Introduction}

A critical test of the $\Lambda$CDM model is its ability to accurately
predict the evolution of the distribution of ``subhaloes'' within dark
matter haloes, or haloes within haloes.  The hierarchical formation process
of CDM haloes by multiple mergers (White \& Rees 1978) leaves behind 
tidally-stripped merger
remnants that survive as bound subhaloes within larger haloes (Ghigna 
\etal 1998).  Subhaloes serve as hosts for visible galaxies
within clusters, groups, or larger galaxies, and so provide a
powerful and observable cosmological probe. In cases where dark matter
subhaloes may have no luminous counterparts, the substructure population 
can be inferred from gravitational lensing studies (\eg Mao \& Schneider 
1998; Metcalf \& Madau 2001; Chiba 2002; Dalal \& Kochanek 2002; Mao \etal 
2004).

CDM models predict the abundance of substructure to be roughly 
self-similar,
independent of halo mass (Klypin \etal 1999;  Moore \etal 1999), and 
to follow a poisson distribution (Kravtsov \etal 2004a). Subhalo
numbers predicted by the $\Lambda$CDM model are reasonably matched by
observations of clusters (Ghigna \etal 2000; Springel \etal 2001; Desai
\etal 2004; see however Diemand, Moore \& Stadel 2004b). However,
observations measure roughly an order of magnitude fewer subhaloes in
Galactic haloes than in clusters (Klypin \etal 1999;  Moore \etal 1999).  
Thus, mass is either more smoothly distributed on small scales than
predicted by $\Lambda$CDM cosmology, or Galactic dark matter subhaloes
are poorly traced by stars. An element of uncertainty in the comparisons
with $\Lambda$CDM model predictions is the possibility of significant
halo to halo variation in the subhalo population that could depend on
host mass, merging history, or environment.  Results from the handfull of
high-resolution simulations on galactic scales to date suggest that such
variance in substructure numbers are significant but much too small to
account for the apparent discrepancy in Galactic subhaloes (\eg Moore
\etal 1999; Klypin \etal 1999; Font \etal 2001).  Even if observed
Galactic dwarfs reside in subhaloes of significantly deeper potential
than inferred from their stellar velocity dispersions and radial extent,
which could allow the most massive satellites to match predictions
(Stoehr \etal 2002; Hayashi \etal 2003; however see also Kazantzidis
\etal 2004 and Willman \etal 2004), then the non-detection of the vast
majority of small subhaloes remains an unsolved problem. However,
semi-analytic work suggests that baryonic physics causes small haloes to
remain starless, indicating that observations may be consistent with the
$\Lambda$CDM model (\eg Bullock, Kravtsov \& Weinberg 2000; Benson \etal
2002ab; Somerville 2002).

By analysing substructure in a large number of dark matter haloes, we can
measure the range of halo to halo variation, and better constrain the
uncertainty in the $\Lambda$CDM subhalo distribution with the improved
statistics. With better numerical resolution and a broad range in host
halo masses, substructure can be used to place cosmological constraints at
new mass scales.  Furthermore, detailed dark matter simulations provide a
theoretical $\Lambda$CDM baseline 
to link subhalo properties with observable galaxy characteristics.
The conditional
luminosity function, which describes the number of galaxies of luminosity 
${\rm L \pm dL/2}$ in hosts of a given mass, can be combined with simulated 
subhalo populations
in order to associate subhalo properties with observable characteristics 
(Yang, Mo 
\& van den Bosch 2003; van den Bosch, Yang \& Mo 2003; see further
Vale \& Ostriker 2004).  This allows one to investigate the impact of
baryonic physics on the galaxy distribution.

Our simulation set is more sensitive to possible dependence of the subhalo
population on host halo mass or redshift than previous works.  
Consequently, we can improve constraints on whether substructure
properties are a function of $M/M_{*}$, where $M_{*}$ is the
characteristic mass of collapsing haloes defined by the scale at which the 
rms linear density
fluctuation equals the threshold for non-linear collapse.  One might
expect that because low-mass haloes were mostly assembled much earlier when
the universe was more dense, are at smaller $M/M_{*}$, and lie where the
power spectrum of mass fluctuations is steeper, that they might have
significantly different subhalo distributions than more massive haloes. If
subhaloes within low-mass and low-redshift haloes have a higher
characteristic density (see \eg Reed \etal 2005 and references therein),
then they should be less subject to tidal stripping and disruption.  
Additionally, if the infall rate on to the host halo (\ie merger rate) is
different from the rate of subhalo destruction, then subhalo numbers will
evolve with redshift.

The angular momentum of a cosmic (sub)halo is crucial to determine the
radial distribution of its eventual stellar component (\eg Mo, Mao \&
White 1998; Verde, Oh \& Jimenez 2002, Van den Bosch \etal 2002)  and its
collapse factor compared to the parent dark matter halo (Stoehr \etal
2002).  If high angular momentum material is systematically stripped from
subhaloes this would further inhibit the formation of stellar discs in
dense environments consistent with observations (\eg Goto \etal 2003).
Also, a high collapse factor of the baryonic component would make the
velocity dispersions of the stars eventually formed lower than that
associated with the parent halo. This would exacerbate the apparent difference
between the observed abundance of galaxy satellites and the predicted
abundance of their host haloes, as discussed in detail in Stoehr \etal
(2002) and Kazantzidis \etal (2004).

The angular momentum of the dark matter component of a subhalo
will be relevant to its stellar properties only if star formation
continues after infall and tidal stripping, which requires that
sufficient baryons remain in the subhalo.  While one generally expects
much of a subhalo's gas to be stripped upon infall, some of that gas
may also be able to cool and form stars.  Complex processes such as
starbursts after infall and redistribution of baryonic angular
momentum under tidal influences may affect the final spin of the baryonic
component.  Simulations predict starbursts will occur in gas-rich
satellites as they are tidally stirred by the central potential (\eg
Mayer \etal 2001), which may explain the continuing star formation in
Galactic satellites.  However, a better understanding of post-infall
star formation is needed to determine whether the angular momentum of
the stellar component is correlated with that of the dark matter
component.

Subhaloes are particularly sensitive to resolution issues (\eg Moore,
Katz \& Lake 1996).  Dark matter haloes, and by extension, subhaloes,
have densities that continually increase toward the halo centre, and so
should be very difficult to disrupt completely unless numerical
discreteness effects artificially lower the central density.  A subhalo
with a numerically softened cusp is more easily disrupted by the global
tidal forces and interactions with other subhaloes that strip away the
outer regions.  Poor spatial resolution and two body relaxation lowers 
central densities, and so may
lead to subhalo destruction, especially near host halo centres. 
Simulated clusters may generally suffer more from resolution issues than 
galaxies because of their later formation epoch, which means that cluster 
particles will have spent more time in low mass haloes (Diemand \etal 2004a). 
Also, subhaloes with
highly eccentric orbits are more likely to be disrupted since they pass
near to the central potential.

In this work, we present the results of substructure analyses of 16
$\Lambda$CDM simulated haloes covering three decades in mass, from dwarfs to
clusters, each with roughly a million particles.  Our sample includes ten
clusters extracted from one cosmological volume (CUBEHI) to study
cosmological variance, and also includes a seven million particle group and
a four million particle cluster.  Some of our haloes are well-resolved to
redshifts of three or higher, allowing investigation of mass or
redshift-dependent trends.  Additionally, we have modelled a cluster in power
law cosmologies where P$\propto$k$^{\rm n}$ to analyse the dependence of the
subhalo distribution on spectral index, $n$.

\section{Numerical Techniques}
\subsection{The Simulations}

We use the parallel K-D (balanced binary) 
Tree (Bentley 1975) gravity solver PKDGRAV (Stadel
2001; see also Wadsley, Stadel \& Quinn 2004) 
to model sixteen dark matter haloes, further described in Reed \etal
(2005); see Table 1.  
The CUBEHI run consists of a cube of 432$^{3}$ particles of uniform
resolution.  The six ``renormalized volume'' runs (\eg Katz \& White
1993; Ghigna \etal 1998) consist of a single halo in a high-resolution
region nested within a lower resolution cosmological volume.  Initial
conditions for these high resolution halo runs are created by first
simulating a low resolution cosmological volume.  Next, a halo of
interest is identified.  To minimise sampling bias,
volume-renormalized haloes are selected by mass with the only
additional constraint that they not lie within close proximity (${\rm 2-3
r_{vir}}$) to a halo of similar or larger mass.  Then, the initial
conditions routine is run again to add small-scale power to a
region made up of high resolution particles that end up within
approximately two virial radii of the halo centre, while preserving
the original large scale random waves.  This process is iterated in
mass resolution increments of a factor of eight until the desired
resolution is achieved.  We have verified that the high-resolution
haloes are free from significant contamination by massive particles.

Our largest halo has seven million particles and most have $\sim 10^{6}$
particles within the virial radius at redshift zero.
Due to the high sensitivity of the subhalo distribution to numerical
resolution effects, we only consider haloes with at least
3.5$\times$10$^{5}$ particles.  We adopt a $\Lambda$CDM cosmology with
$\Omega_m=$ 0.3 and $\Lambda=$ 0.7.  The initial density power spectrum is
normalized such that $\sigma_{\rm 8}$ extrapolated to redshift zero is
1.0, consistent with both the cluster abundance (see \eg Eke, Cole \&
Frenk 1996 and references therein) and the WMAP normalization (\eg Bennett
\etal 2003; Spergel et al. 2003). We use a Hubble constant of $h=$0.7, in
units of 100 km s$^{-1}$ Mpc$^{-1}$, and assume no tilt (i.e. a primordial
spectral index of 1). To set the initial conditions, we use the Bardeen
\etal (1986) transfer function with $\Gamma=\Omega_{\rm m}\times h$.  For
the volume-renormalized runs, we list the effective particle number of the
highest resolution region rather than the actual particle number in Table
1. Numerical parameters are consistent with empirical studies (\eg Moore
\etal 1998; Stadel \etal 2001; Power \etal 2003; Reed \etal 2003).

\begin{table*}
\centering
\caption{Summary of our halo sample at redshift zero.  For volume
renormalized runs, the mass (h$^{-1} \msun$) and particle number of the
central halo is listed.  v$_{\rm cmax, host}$ is the host peak circular
velocity (km s$^{-1}$).  N$_{\rm p,eff}$ is the effective
particle number based on the high-resolution region for renormalized runs.  
v$_{\rm c,lim}$ is the subhalo peak circular velocity (km s$^{-1}$) above
which numerical disruption is insignificant.  }
\begin{tabular}{@{}llllllllll@{}}
   &   M$_{\rm halo}$ & v$_{\rm cmax,host}$ & N$_{\rm 
p,halo}$ 
&  N$_{\rm p,eff}$ & v$_{\rm c,lim}$ & $r_{\rm 
soft}(h^{-1}{\rm kpc}$) & $z_{\rm start}$ & L$_{\rm box}$ (h$^{-1}$Mpc) &\\
\hline
CUBEHI & 0.7-2.1$\times10^{14}$ & 710-1010 & 0.6-1.6$\times$10$^{6}$ &
432$^{3}$ & 50 & 5 & 69 & 50 & 10 clusters\\
GRP1    &  4$\times10^{13}$ & 560 & 7.2$\times10^{6}$ & 1728$^{3}$ & 40 & 
0.625 & 119 & 70 & Fornax mass\\
CL1     &  2.1$\times10^{14}$ & 1020 & 4.6$\times10^{6}$ & 864$^{3}$ & 60 
& 1.25  & 119 & 70 & Cluster\\
GAL1    &  2$\times10^{12}$ & 244 & 0.88$\times10^{6}$ & 2304$^{3}$ & 20 & 
0.469 & 119 & 70 & Milky Way\\
GRP2    &  1.69$\times10^{13}$ & 460 & 0.38$\times10^{6}$ & 864$^{3}$ & 60 
& 1.25 & 119 & 70 & Group\\
DWF1    &  1.88$\times10^{11}$ & 130 & 0.64$\times10^{6}$ & 4608$^{3}$  & 
15 & 0.234 & 119 & 70 & 2 Dwarfs \\
        &  1.93$\times10^{11}$ & 130 & 0.66$\times10^{6}$& & & & &\\
$n=0$ & 1.9$\times10^{14}$ & 1300 & 0.54$\times10^{6}$& 432$^{3}$ & & 2.5 
& 799 & 70 & $P \propto k^{0}$\\
$n=-1$ & 2$\times10^{14}$ & 1110 & 0.55$\times10^{6}$ & 432$^{3}$ & & 2.5 
& 269 & 70 & $P \propto k^{-1}$\\
$n=-2$ & 1.6$\times10^{14}$ & 870 & 0.45$\times10^{6}$ &432$^{3}$ & & 2.5 
& 99 & 70 & $P \propto k^{-2}$\\
$n=-2.7$ & 2.9$\times10^{13}$ & 470 & 0.82$\times10^{5}$ &432$^{3}$ & & 
2.5 & 79 & 70 & $P \propto k^{-2.7}$\\
 \end{tabular}
\end{table*}

\subsection{The Analysis}

The host haloes are defined using the SO algorithm (Lacey \& Cole 1994) with 
a tophat overdensity based on Eke \etal (1996) where the 
$\Lambda$CDM virial overdensity, {\it
$\Delta_{\rm vir}$}, in units of critical density is approximately 100.
Bound subhaloes are identified using SKID (Stadel 2001; 
http://www-hpcc.astro.washington.edu/tools/skid.html), which uses local
density maxima to identify bound mass concentrations independently of
environment.  SKID iteratively ``slides'' particles toward higher densities
until a converged group of particles is found.
The radial extent of each SKID halo is determined by the distribution of bound 
particles, and no predetermined subhalo shape is imposed. 
The maximum circular velocity of each subhalo, $v_{\rm
cmax}$, is calculated from the measured peak of the rotation curve ${\rm
v_c(r) = (GM(<r)/r)^{0.5}}$.  For virialized host haloes, the ratio of
${\rm v_{cmax,host}/v_{c,host}(r_{vir})}$ depends primarily on $M/M_*$, 
and is readily estimated from the halo density profile given in
Reed \etal (2005).  Typical values of ${\rm
v_{cmax,host}/v_{c,host}(r_{vir})}$ are 1.5--1.6 for our dwarfs and
1.05--1.25 for our clusters at redshift zero, declining to smaller values
at higher redshift for each halo.  To test for self-similarity between
haloes, we thus normalize the velocity distribution function (VDF) to the
peak circular velocities of hosts.  We note that a subhalo bound to
another subhalo will sometimes be catalogued as a separate SKID subhalo,
particularly when the chosen linking parameter $\tau$ is small. Our tests
indicate that the velocity distribution function is insensitive to $\tau$
except for the few largest subhaloes. We set ${\rm \tau=2r_{soft}}$ for
the CUBEHI run, and ${\rm \tau=4r_{\rm soft}}$ for all other simulations.  
We consider only subhaloes of 32 or more particles in our analyses,
consistent with Ghigna \etal (1998) and Springel \etal (2001).  
For analyses of subhalo angular momenta, we adopt an empirically motivated
minimum of 144 particles per subhalo; see section 3.2. 
Differential plots of binned quantities use the median value
of each bin, and bin sizes are variable in increments of
$\Delta$log$_{10}=$0.1, but increased when necessary so that no bins are
empty.

\subsection{A Resolution Study and the Radial Distribution of Subhaloes}
\begin{figure}
\begin{center}
\psfig{file=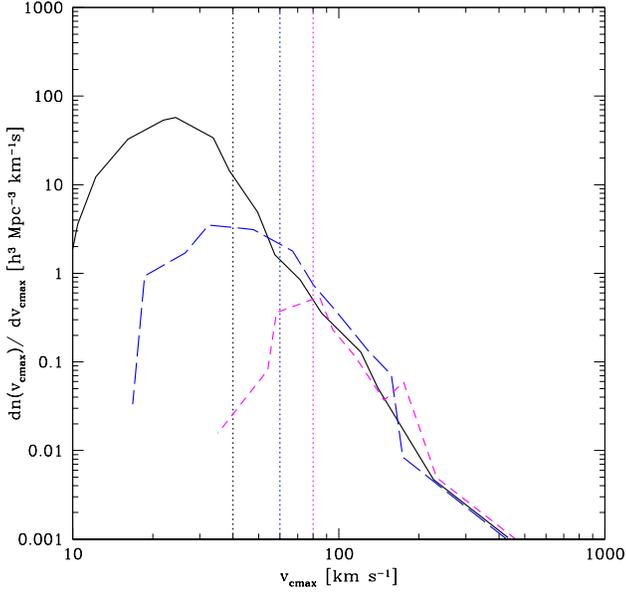, width=\hsize}
\caption{VDF for halo GRP1 (at ${z=0}$) with 7 million particles, 
and for 2 lower
resolutions, with particle mass incrementally increasing by a factor of
eight.  
The circular velocity completeness limit ${\rm v_{c,lim}}$ is marked
for each resolution with a vertical line.}
\label{vclimgrp1}
\end{center}
\end{figure}

For our volume renormalized runs, GAL1, GRP1, and CL1, we have analysed 3,
2, and 1 identical lower resolution versions, respectively, where the mass
resolution is varied by increments of 8$\times$. To find the minimum
subhalo circular velocity down to which our results are complete, ${\rm
v_{c,lim}}$, we plot the subhalo circular velocity distribution
function (VDF) for each simulation, and identify the ${\rm v_{\rm cmax}}$
below which the VDF slope begins to flatten due to incompleteness, as
described in Ghigna \etal (2000). In Fig. \ref{vclimgrp1}, we plot the
resolution criteria for halo GRP1, marking our conservative ${\rm
v_{c,lim}}$ completeness limits. The agreement of the lower resolution
versions shows that this technique is sound.  Less conservative -- but
still apparently sound -- completeness limits for our halo sample would
have yielded roughly ${\rm v_{c,lim} \propto N_{p,halo}^{1/3}}$ for
each of our three haloes with multiple resolutions. Numerical effects on
the subhalo population are likely manifested more strongly for low-redshift 
subhaloes, since they have been subject to potentially disrupting
events for more time.  This implies that measuring ${\rm v_{cmax}}$ at
${z=0}$ should still be valid at higher redshift.  We have verified
that the $z=0$ ${\rm v_{c,lim}}$ for halo GRP1 remains valid at ${z=1}$
where the highest resolution version still has $3.5\times10^{6}$
particles.  
Additionally, we have considered whether the completeness limits,
which were obtained using subhaloes within the entire virial volume,
may change at small radii, where numerical problems are expected to be
stronger.  We find that even when the sample is limited to ${\rm
r<0.3r_{vir}}$, the ${\rm v_{c,lim}}$ derived from the entire virial
volume as described above, remains valid.

\begin{figure} 
\begin{center} 
\psfig{file=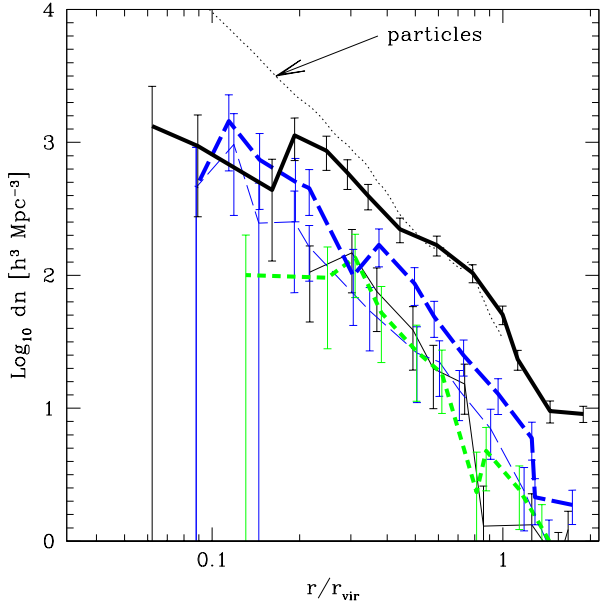, width=\hsize} 
\caption{The ${z=0}$ radial distribution of subhaloes in GRP1 for
each of the 3 different resolutions (solid, long-dashed, short-dashed, 
are the highest, mid, and lowest resolution runs, respectively).
Thin lines have the ${\rm v_{c,lim}}$ resolution limit of the lowest 
resolution run applied.
1$\sigma$ poisson error bars are
shown.  The dotted line is the particle distribution with arbitrary
normalization. } 
\label{grp1dndr} 
\end{center} 
\end{figure}

The radial distribution of the subhalo population, shown for GRP1 in Fig. 
\ref{grp1dndr}, has a slope equal to or shallower than the slope of the 
density profile for each resolution at all radii.  
At very small radii, subhaloes are
highly deficient, possibly due at least in part to artificial disruption by
the tidal forces from the central mass concentration.  When a subhalo
migrates inward via dynamical friction to a radius where its central density
is lower than the local density of the host halo, it will likely be
disrupted (\eg Syer \& White 1998). 
Numerical disruption enhanced by the strong tides near host
centres could affect subhaloes. 
The flattened slope of the radial
subhalo distribution of GRP1 subhaloes relative to the mass profile
interior to roughly 0.2${\rm r_{vir}}$ for our highest resolution, and
interior to 0.3${\rm r_{vir}}$ for our lowest resolution implies that
disruption and/or stripping of subhaloes is important in the halo central 
region. 
The increase in radius of the break in subhalo slope with decreasing 
resolution suggests that numerical disruption, if present, is
worse for lower resolutions.  However,
we caution that the location of the break 
is not well defined because of poisson uncertainties.
Given the uncertainties, it is not possible to reliably separate
spurious from real disruption that may be present in our simulations at
small radii.  Thus, we have no evidence that the central substructure number
density has converged with resolution.
Increasing the mass resolution by a factor
of eight results in roughly a factor of $\sim$2--2.5 more subhaloes at a
given radius beyond roughly 0.3${\rm r_{vir}}$, though there is
substantial noise in this estimate.  
At larger radii, where numerical effects are less important, a subhalo radial
distribution that is shallower than the mass profile is favoured in our data, 
and is also reported by Diemand \etal
(2004b), Gill, Knebe \& Gibson (2004a), Stoehr \etal (2003), 
Gao \etal (2004ab), De Lucia \etal (2003), and Nagai \& Kravtsov (2005).
However, a radial subhalo slope equal to the mass profile slope is not ruled 
out except in the central region where numerical effects may dominate.
Note that if we use only a minimum mass cut without imposing the
circular velocity completeness limit, a more significant anti-bias in
the radial distribution is found (see also Nagai \& Kravtsov (2005)

\section{Results} \subsection{The Circular Velocity Distribution Function
(VDF)} 
\begin{figure}
\begin{center}
\psfig{file=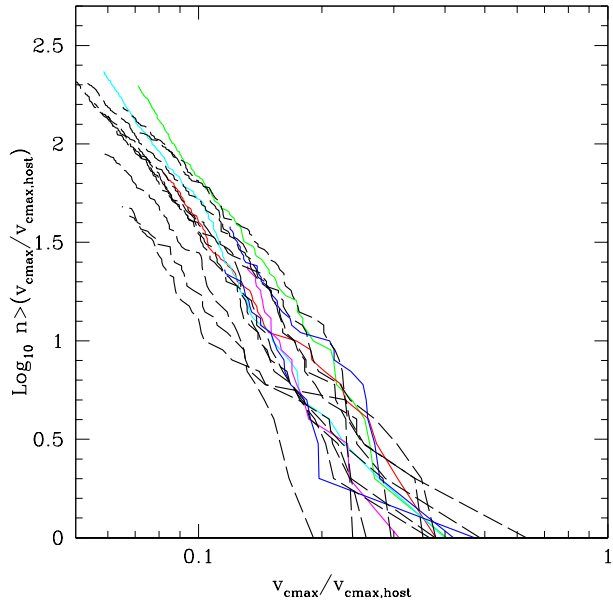, width=\hsize}
\caption{The ${z=0}$ 
cumulative subhalo VDF with the completeness limits 
${\rm v_{c,lim}}$
applied, and considering only subhaloes with 32 or more particles, plotted
as a function of ${\rm v_{cmax}/v_{cmax,host}}$.
Solid lines are renormalized
volumes (green, light blue, magenta, red, and blue for
CL1, GRP1, GRP2, GAL1, and DWF1, respectively).
Dashed lines (black) are the ten clusters in the CUBEHI simulation.
}
\label{vdfcum}
\end{center}
\end{figure}

\begin{figure}
\begin{center}
\psfig{file=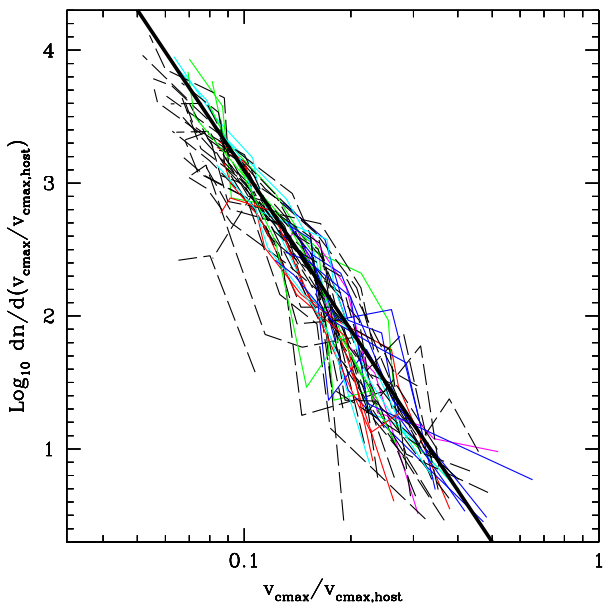, width=\hsize}
\caption{Normalized subhalo VDF for all haloes with
${\rm N_{p}>3.5\times10^{5}}$ at
redshifts 0, 0.5, 1, and 2,
normalized to ${\rm v_{cmax, host}}$
and a virial volume of unity.
Solid lines (colours as in Fig. \ref{vdfcum})
are renormalized
volumes, dashed lines (black) are the ten clusters in the CUBEHI 
simulation.
Heavy solid line corresponds to
${\rm dn/dv = {1 \over 8} (v_{cmax}/v_{cmax,host})^{-4}}$.
}
\label{vdfvirall}
\end{center}
\end{figure}

\begin{figure}
\begin{center}
\psfig{file=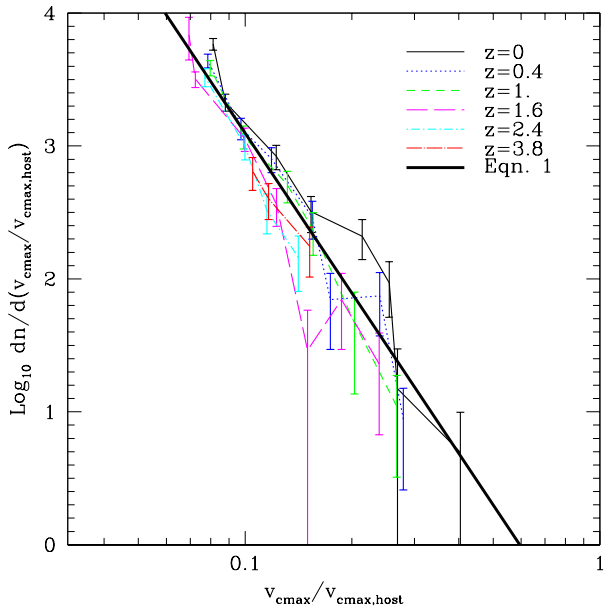, width=\hsize}
\caption{Evolution of the normalized subhalo VDF for halo GRP1.
Subhalo circular velocities are normalized to ${\rm v_{cmax, host}}$
and a virial volume of unity.}
\label{vdfvirgrp1z}
\end{center}
\end{figure}

In Fig. \ref{vdfcum}, we plot the cumulative VDF for all haloes at
redshift zero. Our completeness and resolution limits exclude the majority
of subhaloes identified by SKID, leaving of the order of 100 subhaloes in most
hosts. Fig. \ref{vdfvirall} shows the differential VDF for our entire set
of subhaloes at $0\leq$z$\leq$2 for hosts of more than $3.5\times10^{5}$
particles.  The subhalo VDF is well-approximated by a power law with slope
and normalization given by
\begin{equation} {\rm dn/dv = {1 \over
8}({v_{cmax} \over v_{cmax,host}})^{-4}}, 
\label{vdvfit} \end{equation} 
over the range of approximately 0.07${\rm v_{cmax,host}}$ to 0.4${\rm
v_{cmax,host}}$ with halo to halo scatter of a factor of roughly two to
four.  When we individually consider the VDF for hosts of different mass,
we find no evidence of mass dependence on the VDF amplitude or slope in
our data, and similarly we detect
evidence of weak or zero redshift dependence (see below).
This implies that the number density of subhaloes is approximately 
self-similar, independent of the mass and redshift of the host halo.
Here we caution that a larger halo sample would be needed to identify
any weak trends (\eg an increase in substructure abundance with halo
mass, reported by Gao \etal 2004a) as they would be masked by the
large halo to halo scatter.
The farthest low
VDF outlier halo is the largest cluster in the CUBEHI simulation at 
${z=1}$,
which has approximately $4.5\times10^{5}$ particles at the time. In Fig.
\ref{vdfvirgrp1z}, we present the VDF of our highest resolution halo to
redshift ${z\simeq 4}$.  There is little or no evolution in VDF slope or 
normalization, although there are
fewer large subhaloes at high redshift, as seen by the lack of
data points in the VDF beyond large ${\rm v_{cmax}/v_{cmax,host}}$ in 
high-redshift hosts.  Thus, large subaloes are deficient until lower redshifts
when they either infall on to the host or are formed as merger products of
existing subhaloes.  

\begin{figure}
\begin{center}
\psfig{file=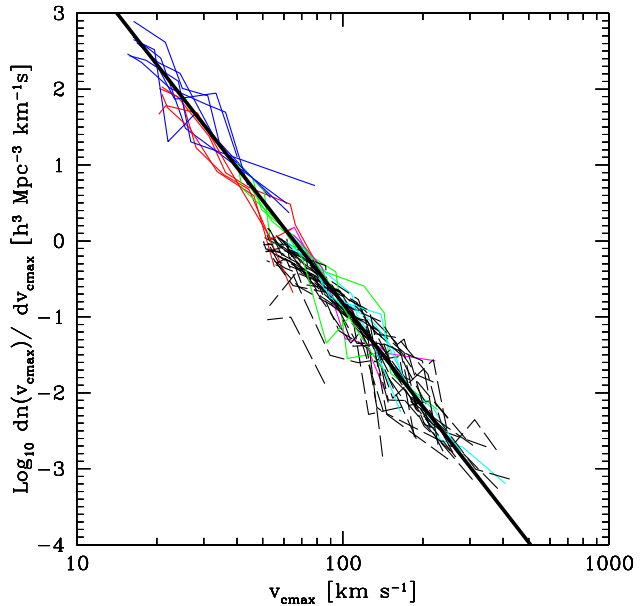, width=\hsize}
\caption{Differential subhalo VDF for all haloes with $N_{\rm p} > 
3.5\times10^{5}$ at
redshifts 0, 0.5, 1, and 2.  Solid lines (colours as in Fig. \ref{vdfcum})
are renormalized
volumes, dashed lines (black) are the ten clusters in the CUBEHI simulation.
The heavy solid line denotes ${\rm dn/dv=0.5\times10^{8}v_{\rm 
cmax}^{-4.5}}$ (Eq. \ref{vkmsfit}).  
Volume units are comoving.}
\label{vdfcoall}
\end{center}
\end{figure}

\begin{figure}
\begin{center}
\psfig{file=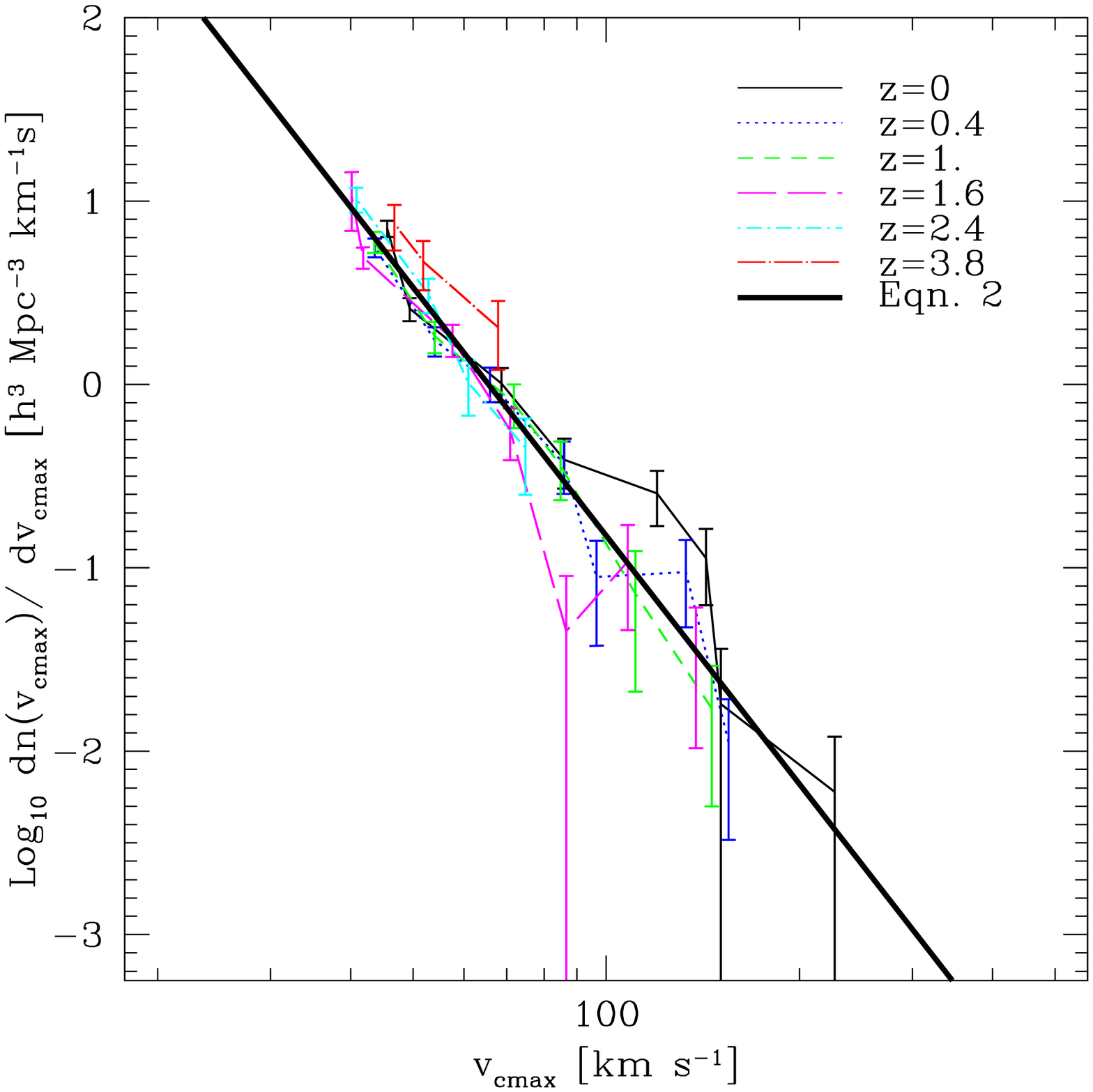, width=\hsize}
\caption{Subhalo VDF in comoving coordinates
for the group GRP1, which still has over half a million
particles at $z=3.8$.  1 $\sigma$ poisson error bars are shown.}
\label{vdfgrp1zco}
\end{center}
\end{figure}

In Fig. \ref{vdfcoall}, we present the non-normalized differential VDF
for our entire set of subhaloes, which displays little or no redshift
evolution in comoving space. The subhalo VDF is fit by 
\begin{equation}
{\rm dn/dv = 1.5\times10^{8}v_{\rm cmax}^{-4.5} (h^3 Mpc^{-3} km^{-1}s)},
\label{vkmsfit} 
\end{equation} 
for subhaloes of 20--300 ${\rm km s^{-1}}$ in hosts of 10$^{11}$ to
10$^{14}$ $\msun$ and redshifts from zero to two, with scatter of a factor
of $\sim 2-4$.  This slope of this non-normalized VDF is steeper because
lower mass galaxies in the sample have a larger ratio of ${\rm
v_{cmax,host}/v_{c,host}(r_{vir})}$. The dependence on ${\rm
v_{cmax,host}/v_{c,host}(r_{vir})}$ implies that this non-normalized VDF
is unlikely to hold universally. Fig. \ref{vdfgrp1zco} shows that the
non-normalized subhalo VDF for GRP1 has weak or no evolution to ${z \simeq
4}$.

\begin{figure}
\begin{center}
\psfig{file=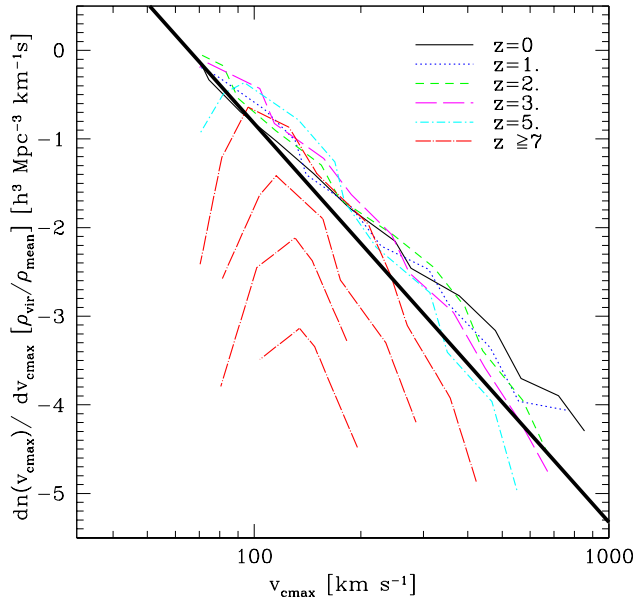, width=\hsize}
\caption{VDF for $FOF$ haloes in the CUBEHI simulation plotted for ${0 
\leq z \leq 15}$.  Heavy solid line 
is the fit to the subhalo population given by Eq. 2 (non-normalized).  For 
reference, we normalize the FOF VDF to 
the virial overdensity as a function of redshift by multiplying by 
a factor of ${\rm \rho_{virial}/\bar{\rho}}$.}  
\label{vdffof}
\end{center}
\end{figure}

We also calculate the VDF of {\it friends-of-friends} (FOF, Press \& Davis
1982; Davis \etal 1985)  haloes, from which the population of subhaloes
must have descended. The FOF algorithm identifies virialized
haloes that are not part of larger virialized objects.  In Fig.
\ref{vdffof}, we plot the VDF for all FOF haloes in the CUBEHI simulation
adopting a linking length of 0.2 times the mean particle separation. We
have normalized the FOF VDF by multiplying it by the mean virial
overdensity, ${\rm \rho_{virial}/\bar{\rho}}$, at each redshift.  
Differences between the
subhalo and halo VDFs are generally smaller than a factor of three until
${z \simgt 7}$, in agreement with ${z=0}$ results by Gao \etal
(2004a).  At ${z\simgt5}$, the FOF VDF drops rapidly with increasing
redshift, presumably due to the fact that it samples the steep drop-off
regime of the mass function at high redshift.  Higher redshift simulations
are needed to test whether the lower FOF VDF at ${z \simgt 7}$ results
in small subhalo numbers for extremely high-redshift hosts.

\subsection{Subhalo Angular Momenta}
\begin{figure}
\begin{center}
\psfig{file=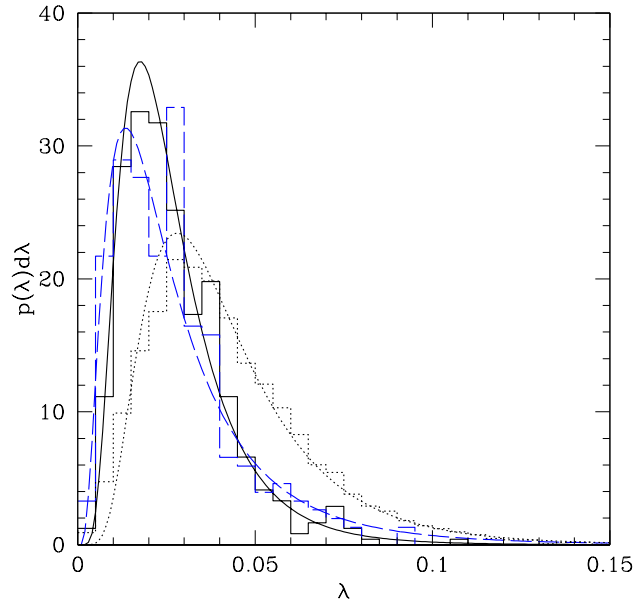, width=\hsize}
\caption{Histograms show the ${z=0}$ 
distribution of spin subhalo parameters within
halo GRP1 (solid) and halo CL1 (dashed) for subhaloes with 144 or more 
particles.  Spins of SO field haloes from the CUBEHI volume are shown by 
dotted lines.  Curves are fits to the lognormal distribution described in 
text.}
\label{spinlog}
\end{center}
\end{figure}

We analyse the subhalo angular momenta given by the spin parameter,
$\lambda = LE^{1/2}/GM^{1/2}$ (Peebles 1969), for subhaloes at redshift zero 
in our two 
highest resolution simulations, GRP1 and CL1.  Angular momentum $L$ is 
calculated with respect to the centre of mass of the subhalo.  Energy 
$E=E_{\rm kinetic} + E_{\rm potential}$ is summed over all subhalo particles.
Each central subhalo and subhaloes with more than 50,000 particles (for 
computational speed) are excluded.  
In our simulations, median subhalo $\lambda$ increases with decreasing 
particle number once below $\sim$100 particles, indicating an upward bias 
for poorly resolved haloes.
We thus limit our spin analyses to subhaloes containing
144 or more particles, which results in no dependence of median spin on
subhalo mass.  Several studies have found 
the lognormal function to be a good description of halo spins
\begin{equation}
{\rm p(\lambda){\rm d} \lambda = {1 \over \sigma_{\lambda} \sqrt{2 \pi}}
\exp\biggl(- {{\rm ln}^2(\lambda/\lambda_0) \over 2
  \sigma^2_{\lambda}}\biggr) {{\rm d} \lambda \over
  \lambda} }
\label{lognorm}
\end{equation}
(Barnes \& Efstathiou 1987; Ryden 1988; Cole \& Lacey 
1996; Warren \etal 1992; Gardner 2001; van den Bosch \etal 2002; Vitvitska
\etal 2002; Colin \etal 2004;  Peirani, 
Mohayaee \& Pacheco 2004; Aubert, Pichon \& Colombi 2004)) where 
$\lambda_0$ and 
$\sigma_{\lambda}$ are fit parameters.
Fig. \ref{spinlog} compares the histogram of subhalo spins with the 
best-fitting $\chi^{2}$ lognormal function given by ($\lambda_0, 
\sigma_{\lambda}$) $=$ (0.0235, 0.54) and (0.0238, 0.73) for halo GRP1 and 
CL1, respectively.  Median and arithmetic average values of ${\rm 
\lambda_{med}=0.024}$ and ${\rm \lambda_{avg}=0.027}$ for GRP1 subhaloes, 
and ${\rm \lambda_{med}=0.024}$ and ${\rm \lambda_{avg}=0.026}$ for CL1 
subhaloes.  
Subhalo spins are significantly smaller than spins of a sample containing
${\rm 1.5 \times 10^{4}}$ field haloes selected from the CUBEHI volume using 
the spherical overdensity (SO) algorithm, which has ${\rm
\lambda_{med}=0.037}$ and ($\lambda_0, \sigma_{\lambda}$) $=$ (0.037, 0.57)
over the mass range of ${\rm 1.9 \times 10^{10}-6.5 \times 10^{12}}$ 
(144--50000) particles.  
Within this field halo sample and
within the subhalo samples, spins have no significant mass dependence, so 
comparisons between differing mass scales of different simulations should be 
valid.
Our field angular momenta are consistent with a number
of recent studies that find $\lambda_0=0.035-0.046$ for virialized
$\Lambda$CDM haloes (Bullock \etal 2001; van den Bosch \etal 2002;
Vitvitska \etal 2002; Colin \etal 2004; Peirani \etal 
2004; Aubert \etal 2004).  Note that some of these studies 
use a slightly different definition of spin introduced by Bullock \etal 
(2001) that has some dependence on the density profile.

Given that our subhaloes are in high-density environments, it is likely
that their spins are lowered by the removal of high angular momentum
material, which should be most vulnerable to tidal stripping.  
We verify that stripping of outer material has the potential to lower spins 
by the required amounts by measuring
spins for the central regions of SO haloes in the CUBEHI volume.  Here we find 
that the central SO region containing 15$\times$ the usual virial overdensity
has ${\rm \lambda_{med}=0.025}$, which is similar to ${\rm \lambda_{med}}$ for
subhaloes.  This overdensity generally contains the central $\sim$15--30$\%$ 
of the SO mass.  We have also determined that the SO haloes whose central 
material  have the lowest spins relative to the surrounding halo material are 
the high-spin SO haloes that comprise the tail of the lognormal distribution.  
This suggests that subhaloes with initially 
high spins are likely to be the largest contributors to the removal of 
subhalo angular momentum.  Interestingly, the central regions of
low-spin SO haloes (${\rm \lambda << \lambda_{med}}$) generally have higher 
spin than the surrounding halo material.  

An alternative explanation for the low spins of subhaloes may be
related to the fact that (sub)halo mass growth is generally halted
upon infall into a larger halo.  For the case of field haloes, spins
decrease over time in the absence of major mergers, particularly in
the period immediately following a major merger event (\eg Vitvitska
\etal 2002; Peirani \etal 2004).  If a similar decrease in spin occurs
in the high density environments within haloes, then it could
contribute to the low subhalo spins.  More study is needed to
determine whether this mechanism for lowering spin is effective for
subhaloes.

\begin{figure}
\begin{center}
\psfig{file=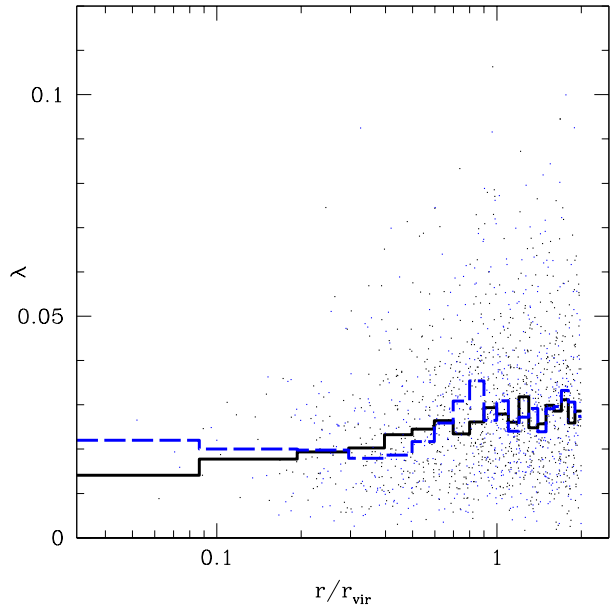, width=\hsize}
\caption{Radial dependence of spin parameter at ${z=0}$.  
Solid (dashed) histogram is 
median of subhalo spins for host GRP1 (CL1)
}
\label{spinvsr}
\end{center}
\end{figure}

The radial
dependence of subhalo spins shown in Fig. \ref{spinvsr} shows that local
environment directly affects subhalo spin. Median subhalo spin decreases
with decreasing radius from $\lambda\simeq0.025-0.03$ near ${\rm r_{vir}}$ to
$\lambda\simeq0.02$ at 0.3${\rm r_{vir}}$, where tidal effects are 
greater.  The transition in spins from subhaloes to field haloes 
at ${\rm r_{vir}}$ is smooth, and spins of SKID haloes found between
${\rm r_{vir}}$ and 2${\rm r_{vir}}$ have weak or no further radial dependence.
Note that the median SKID halo spin ${\rm
\lambda_{med}\simeq0.03}$ between ${\rm r_{vir}}$ and 2${\rm r_{vir}}$
is somewhat smaller than for SO haloes selected from the uniform volume.
We have tested that when the field SO haloes of the large volume (CUBEHI) 
are instead selected with SKID, their spins are smaller than with SO, but 
only by $5-10 \%$, which
is not enough to explain the small spins found in the high-density regions.
This suggests that field haloes in the high-density environments just 
outside of larger virialized haloes have 
spins that are $\sim$10$\%$ lower than for the global population.  Spins in 
high-density regions could be reduced due to a contribution from subhaloes 
whose orbits have previously taken them within the virial radius (Gill, Knebe 
\& Gibson 2005), causing stripping of high angular momentum material.
Additionally, tidal interactions with the massive neighbouring halo 
(Gnedin 2003; Kravtsov, Gnedin \& Klypin 2004b) may also have some impact on 
spin.  This relatively weak environmental dependence on field halo spins is 
still consistent with Lemson \& Kauffmann (1999), who found no
difference in spins of virialized haloes in mild over(under)-densities.

The radial trend in spins
could be observable if reliable spins can be estimated for a large sample
of cluster galaxies or satellites in external systems, 
and if spin of baryonic material can be used as a tracer of subhalo spin.  
Galaxies that form from low spin material should have larger collapse factors.
If star formation occurs after subhalo angular momentum has been lowered,
then galaxies nearer to host centres may thus have smaller radial extent
than galaxies near or
outside the virial radius, although complex baryonic processes related to
star formation may dominate over any potential spin-induced trend in stellar 
distribution 
(see Stoehr \etal 2002; Hayashi \etal 2003; Kazantzidis \etal 2004).
The central  
galaxy would likely deviate from this trend since it undergoes late mergers, 
which have been shown to increase halo angular momentum (Gardner 2001; 
Vitvitska \etal 2002).

A radial trend in subhalo spins is unlikely to be easily
observable unless the stellar component of a galaxy is assembled
subsequent to infall and stripping of high spin material.  
If gas infall, star formation, and disc growth all
cease soon after infall, then spins of
galaxies hosted by subhaloes would only reflect the spin of the original
dark matter halo.  If however, sufficient gas remains after
accretion into another dark matter halo that there is some continuing
star formation (as for Galactic satellites), then the stellar and
gaseous distribution may have a smaller spin than that of the dark
matter host prior to infall.  Numerical simulations are needed to
determine the extent that subhalo baryonic spins reflect the spins of
their host dark matter subhaloes.

\subsection{Subhalo Mass Function} 
\begin{figure}
\begin{center}
\psfig{file=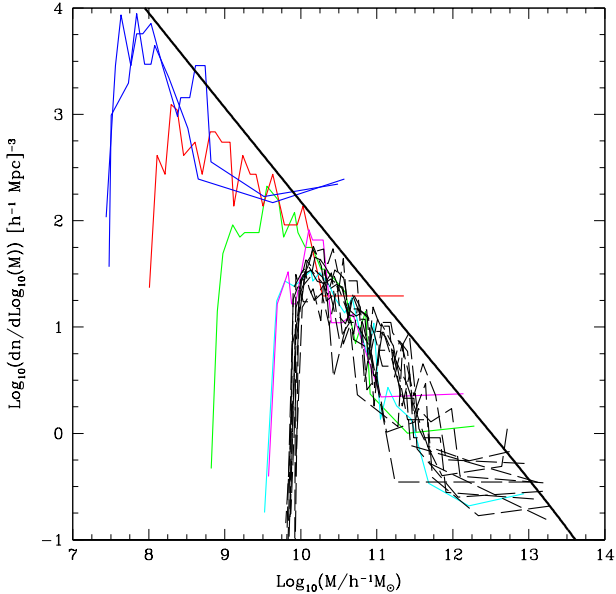, width=\hsize}
\caption{The mass function of the subhalo population for
each of our haloes at $z=0$.  Heavy dark line is the Sheth and Tormen (1999)
prediction for virialized haloes, which closely matches the low 
redshift FOF 
mass function, and is normalized by a simple 
factor equal to the virial overdensity 
(${\rm \rho_{virial}/\bar{\rho}}$) for reference.}
\label{ndmall}
\end{center}
\end{figure}

\begin{figure}
\begin{center}
\psfig{file=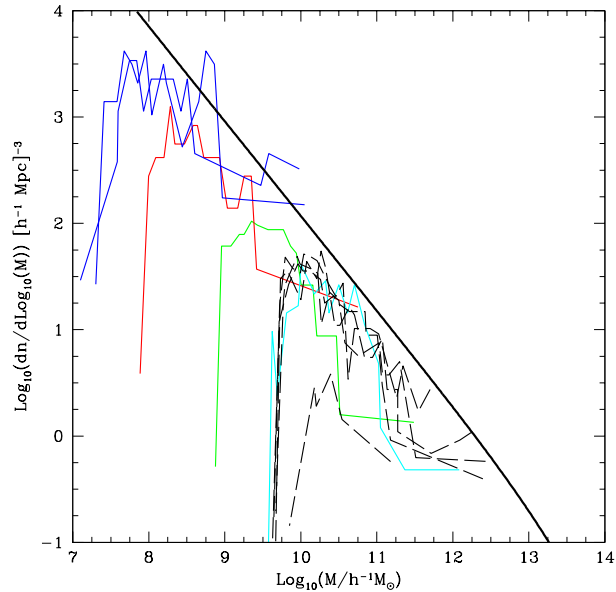, width=\hsize}
\caption{Subhalo mass function, as in Fig. \ref{ndmall}, except at $z=1$.}
\label{ndmallz1}
\end{center}
\end{figure}

We plot the subhalo mass function at
${z=0}$ (Fig.  \ref{ndmall}) and at ${z=1}$ (Fig. \ref{ndmallz1}). 
The steep
drop on the low mass end for each halo is due to exclusion of haloes with
${\rm v_{cmax}<v_{c,lim}}$ for each simulation.  The subhalo mass
function is independent of halo mass, a result also seen in De Lucia \etal
2004, although a weak mass-dependence such as that reported by 
Gao \etal 2004a is not ruled out by our data.  
The Sheth and Tormen (1999) function normalized by a factor
equal to the virial overdensity, is plotted for reference.  The Sheth and
Tormen function, a modification of the Press and Schecter (1974)
formalism, is an excellent match to the CUBEHI FOF mass function at low
redshifts (Reed \etal 2003).  The factor of approximately two to three
offset
between the subhalo mass function and the FOF mass function is independent
of mass and redshift, which implies that the stripping efficiency of
subhaloes is largely mass and redshift independent, though infall timing and 
evolution of the global mass function may also affect the subhalo mass 
function.

\subsection{Radial Distribution and Subhalo Orbits}

\begin{figure}
\begin{center}
\psfig{file=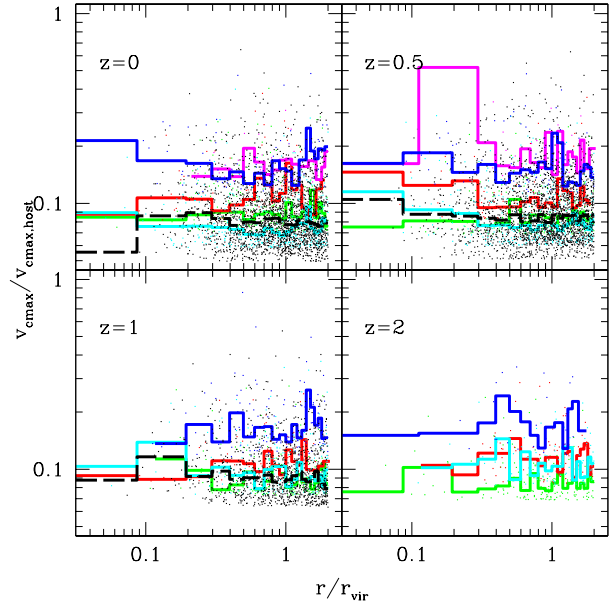, width=\hsize}
\caption{The ${\rm v_{cmax}/v_{cmax,host}}$ distribution of subhaloes plotted 
against radial position at redshifts 0 (top left), 0.5 (top right), 1 
(bottom left), and 2 (bottom right).  Individual histograms represent the 
median
${\rm v_{cmax}/v_{cmax,host}}$ from each simulation with colours as in Fig. 
\ref{vdfcum} (\eg dashed (black) histogram is an average of 10 CUBEHI 
clusters). 
Coloured points are from the normalized
runs with same colours as before, and black points are from the
CUBEHI run.  The subhalo associated with the potential center 
is excluded from each halo.
Subhaloes from hosts with less than $3.5\times10^{5}$
particles are again excluded.
}
\label{vcvsrpoints}
\end{center}
\end{figure}

\begin{figure}
\begin{center}
\psfig{file=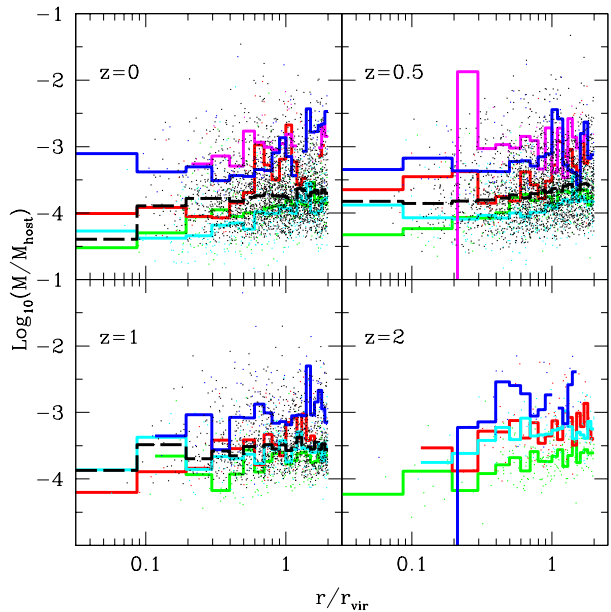, width=\hsize}
\caption{The ${\rm M/M_{host}}$ distribution of subhaloes plotted against
radial position at redshifts 0 (top left), 0.5 (top right), 1 (bottom
left), and 2 (bottom right).  Individual histograms represent the median
${\rm M/M_{host}}$ from each simulation as in Fig. \ref{vcvsrpoints}.
}
\label{momvsrpoints}
\end{center}
\end{figure}

In Fig. \ref{vcvsrpoints}, we show the distribution of subhalo 
${\rm v_{cmax}}$ versus radial position for a range of redshifts.  There are 
no clear radial trends in ${\rm v_{cmax}}$ for ${z\leq2}$. However, Fig.
\ref{momvsrpoints} shows that subhaloes near the centres of their hosts
tend to have lower median masses than subhaloes at larger radii, a trend
consistent with \eg Taffoni \etal (2003) (see also De Lucia \etal 2004;  
Taylor \& Babul 2004; Kravtsov \etal 2004b; Gao \etal 2004a), 
indicative of tidal stripping near
halo centres (\eg Tormen, Diaferio \& Syer 1998). This suggests that the
morphology-radius relation seen in clusters (\eg Whitmore \& Gilmore 1991)
cannot be explained by a ${\rm v_{cmax}}$-radius relation.  We note that
the region where the median mass is smallest (less than $\sim$0.3r$_{\rm
vir}$ is also the least numerically robust, discussed in section 2.3.

\begin{figure}
\begin{center}
\psfig{file= 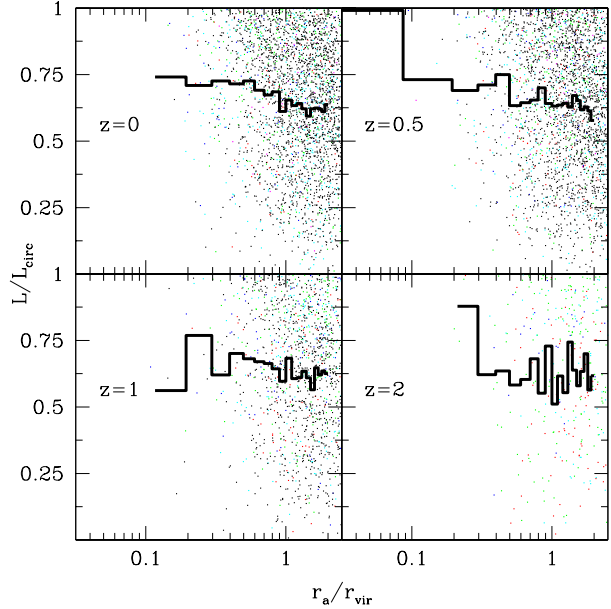, width=\hsize}
\caption{The circularity of subhalo orbits, ${\rm L/L_{circ}}$ vs. 
${\rm r_a/r_{vir}}$, for the
same haloes and redshifts as in Fig. \ref{vcvsrpoints}--\ref{momvsrpoints}. 
${\rm L_{circ}}$ is the orbital angular momentum of
a subhalo on a circular orbit of a given orbital energy.  ${\rm r_a}$ is the
subhalo apocentre.  Histogram represents the average ${\rm L/L_{circ}}$.
}  
\label{epsvsra}
\end{center}
\end{figure}

\begin{figure}
\begin{center}
\psfig{file=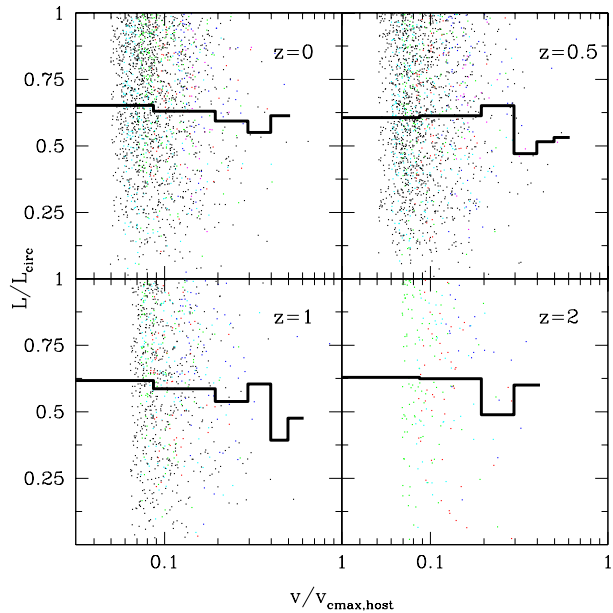, width=\hsize}
\caption{The circularity of subhalo orbits, ${\rm L/L_{\rm circ}}$ vs. 
${\rm (v_{cmax}/v_{cmax,host})}$, for the
same redshifts as in Fig. \ref{vcvsrpoints}--\ref{epsvsra}, here limited to 
subhaloes within ${\rm r_{vir}}$.  Average circularity given by 
histogram.
}
\label{epsvsvpoints}
\end{center}
\end{figure}

In Fig. \ref{epsvsra}, we show the circularity of subhalo orbits, ${\rm
L/L_{circ}}$, which is the angular momentum that a subhalo with a given
orbital energy would have if it were on a circular orbit. 
Orbits are calculated from a snapshot position and velocity of
each substructure applied to a static spherical approximation of the host
halo potential, which is computed from the mass profile as in Ghigna \etal
(1998). Our subhaloes have mean circularities between 0.6 to 0.7 where
sampling is high, with a redshift zero mean ${\rm L/L_{\rm circ}=0.64}$
for $r<{\rm r_{vir}}$.  
The subhalo circularity increases weakly at small 
radii for lower redshifts, suggesting circularisation of orbits over time as 
seen in simulations by (Gill \etal 2004).
Also, there are a larger number of subhaloes in the
nearly circular orbits than in the most radial orbits, especially at low
redshift.  Both of these trends may be a result of disruption or heavy 
stripping of subhaloes on extremely radial orbits since they pass nearer the 
cluster central
potential.  This effect would be greatest for subhaloes with small apocentres.
In Fig. \ref{epsvsvpoints}, we present the circularity as a
function of ${\rm v_{cmax}/v_{cmax,host}}$.  Again, there are no strong trends,
except that small subhaloes with very radial orbits appear to be
relatively deficient, especially at low redshift.

\begin{figure}
\begin{center}
\psfig{file=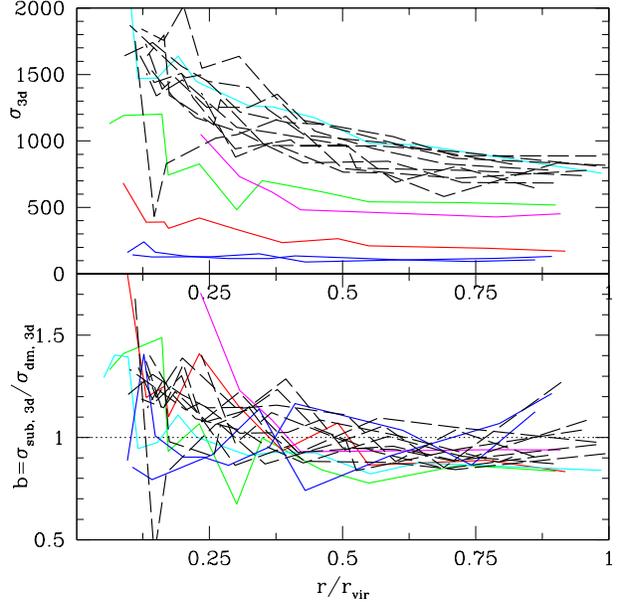, width=\hsize}
\caption{Top: $\sigma_{\rm 3D}$ vs. radius at $z=0$ for each halo.  
Bottom: Subhalo velocity bias, ${\rm b = \sigma_{sub, 
3D}/\sigma_{3D, dm}}$.  Line types and colours are the same as Fig. 
\ref{vdfcum}.
}
\label{sigvsrtile}
\end{center}
\end{figure}

\begin{figure}
\begin{center}
\psfig{file=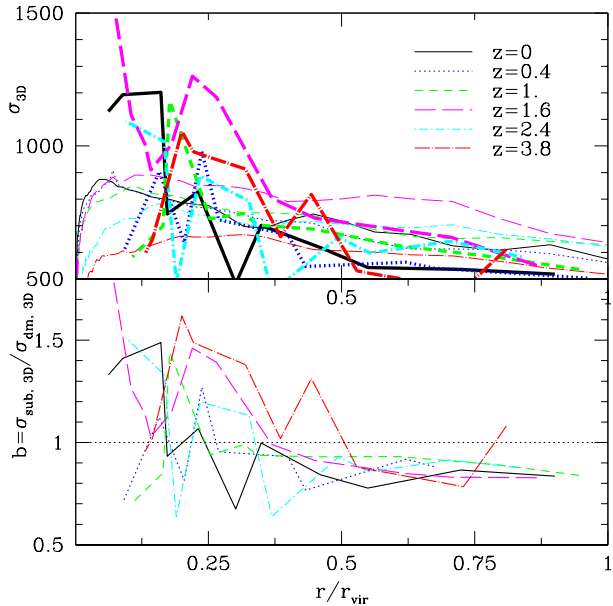, width=\hsize}
\caption{Top: Evolution of velocity dispersions for 
the 7 million particle group GRP1.
Thick lines are $\sigma_{\rm 3D}$ of the subhalo distribution.  
Thin lines are $\sigma_{\rm 3D}$ for particle distribution. Bottom:  
Subhalo velocity bias for same redshifts.
}
\label{sigvsrparts}
\end{center}
\end{figure}

Fig. \ref{sigvsrtile} shows the three dimensional velocity dispersion of
subhaloes, $\sigma_{\rm 3D}$, as a function of radius at redshift 0.  The
velocity dispersion is flat or slowly decreasing with radius at ${\rm \geq
0.5r_{vir}}$. At smaller radii, $\sigma_{\rm 3D}$ generally increases
moderately toward the centre, reaching less than 1.5--2 $\sigma_{\rm
3D}(r_{\rm vir})$.  Subhalo orbital motions suggest a velocity bias
b=${\rm \sigma_{sub, 3D}/\sigma_{3D, dm}}$ (bottom panel of Fig.  
\ref{sigvsrtile}) that increases with decreasing radius, reaching
$\sim 1.2$ by ${\rm \sim0.2r_{vir}}$, similar to that found in \eg 
Colin, Klypin, \& Kravtsov (2000), Ghigna \etal (2000), and 
Diemand \etal (2004b).  
Diemand \etal attribute both 
their central velocity bias and their central spatial anti-bias to tidal 
destruction of slow moving subhaloes near the cluster centre.  
Note that in a dynamically relaxed system, a spatial anti-bias automatically
results in a positive central velocity bias (see \eg van den Bosch 2004).
At radii ${\rm \simgt0.5r_{vir}}$,
there is also a hint of a weak velocity anti-bias of $\sim10\%$ in many
haloes, although subhalo velocities are consistent with no large radius
anti-bias when the range of halo to halo scatter and uncertainty due to
small numbers of subhaloes is considered. In Fig. \ref{sigvsrparts}, we
plot the subhalo $\sigma_{\rm 3D}$ for GRP1 to redshift 4 against 
$\sigma_{\rm 3D}$ for particles.  The subhalo velocity bias and 
$\sigma_{3D}$ are each consistent with no redshift evolution.

\subsection{Power Law Cosmologies} 
\begin{figure}
\begin{center}
\psfig{file=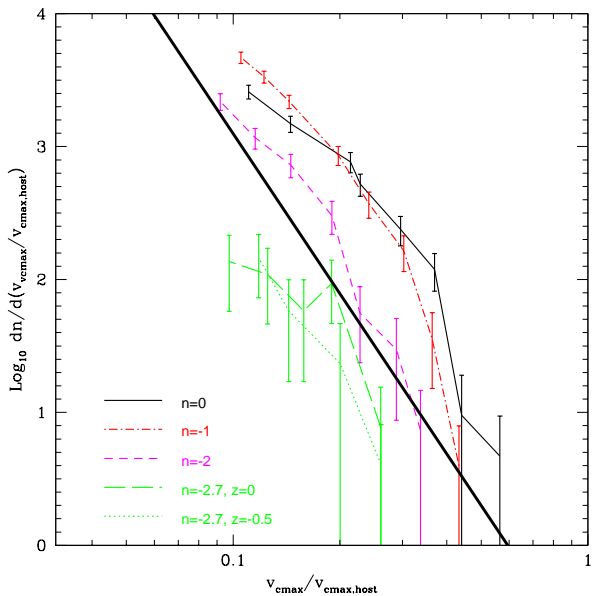, width=\hsize}
\caption{Subhalo VDF for our cluster with initial
power spectrum given by $P \propto k^{n}$.  Heavy solid line is a fit to 
$\Lambda$CDM haloes.
}
\label{plawsub}
\end{center}
\end{figure}

In order to examine the effects of the
power spectral slope index, $n$, we have simulated a single renormalized
volume cluster with a range of values for $n$.  Here the initial density
fluctuation power spectrum is given by $P \propto k^{n}$, normalized to
$\sigma_{\rm 8}=1.0$ as in the $\Lambda$CDM simulations.  We plot the
subhalo VDF for the cluster with initial conditions given by $n=$0, -1,
-2, and -2.7 in Fig. \ref{plawsub}. There is a clear and strong trend
that steeper power spectra have less substructure.  This is a direct
result of the fact that the shallower spectra have more small scale power
relative to large scale power than the steeper spectra. Additionally,
subhaloes form earlier in cosmologies with flatter power spectra, and
have higher characteristic densities and steeper density profiles (\eg 
Reed \etal 2005), making them less vulnerable to disruption.

\section{Conclusions}

(1) {\it A ``universal'' subhalo VDF independent of host mass and
redshift:} The subhalo follows the self-similar relation: ${\rm
dn/dv=(1/8)(v_{cmax}/v_{cmax,host})^{-4}}$ with a factor of two to four
scatter.  

(2) {\it Subhalo spins decrease toward the host centre.} The radial
dependence of subhalo spins is likely explained by the increased
vulnerability of high-spin material to tidal stripping and disruption.  
Galaxies that form from low angular momentum subhaloes in high-density
regions after some stripping has occurred would have a larger baryonic 
collapse factor.
If we assume the size of the stellar distribution is proportional to 
$\lambda$ (as in \eg Kravtsov \etal 2004b), then stellar systems could be 
$\sim$50$\%$ smaller if they form in heavily stripped central subhaloes.
This would require that star formation continue after the subhalo 
has been stripped of high angular momentum material, as discussed previously. 
The circular velocity of the final stellar distribution
could be lowered by a similar amount if the stars collapse to a small
radius, assuming a central subhalo density slope near r$^{-1}$ as found in 
simulations by Kazantzidis \etal (2004).  
Stoehr \etal (2002) proposed that the the apparent deficit of local group
satellites could be explained if satellites actually reside in large 
subhaloes (further supported by Hayashi \etal 2003), which would allow the 
VDF of large Local Group satellites to match $\Lambda$CDM predictions.
The small radii of the stellar distribution
would give them low stellar velocity dispersion relative to 
${\rm v_{cmax}}$ of their dark matter subhalo hosts.
Baryonic spins of $\lambda < 0.01$ would be required for
the gas to collapse to the required $\sim$1$\%$ of the 
pre-stripped virial radius
of the dark subhalo (Kazantzidis \etal 2004) needed to match Local Group
satellites with $\Lambda$CDM predictions.
However, only $\sim$ 5$\%$ of our subhaloes have spins of 
$\lambda \simlt 0.01$, and 90$\%$ have $\lambda>0.014$, so such a solution
appears unlikely if the steep central substructure density profiles found 
by Kazantzidis \etal are correct.

(3) {\it Evidence for redshift dependence of the VDF:} The shape and
amplitude of the subhalo VDF has little or no trend with mass or redshift.
However, fewer subhaloes lie at large ${\rm v_{cmax}/v_{cmax,host}}$ in
high-redshift hosts.  This implies that dark matter haloes are not
populated with large subhaloes until lower redshifts.

(4) {\it Could the VDF be a function of $M/M_{*}$?} Though our data is
consistent with only a weak or no trend of the VDF on host mass and
redshift, 
a trend of less substructure in large $M/M_{*}$ haloes would not be 
surprising, although it remains somewhat speculative until simulations are
able to probe a larger mass and redshift range.
The subhalo population
approximately follows some fraction of the universal halo population.  Thus,
at large $M/M_{*}$, where the mass function is steep, the subhalo mas function
should also be steep. 
The reason we see no such trend among redshift zero haloes in our sample
may be because $M/M_*$ for our simulated haloes is never much larger 
than unity, below which the slope of the field mass function 
has little mass dependence.

A dependence of substructure internal density profiles on $M/M_*$ could also
cause a trend for large $M/M_*$ hosts to have decreased substructure.
Virialized haloes have density profiles that become less centrally 
concentrated with increasing $M/M_*$ (see \eg Reed \etal 2005), so subhalo
density profiles are likely to have a similar dependence.
The lower central densities of low $M/M_{*}$ subhaloes should 
make them more vulnerable to stripping and major disruption during and after
virialization, even though their hosts also have lower central densities.
The potential decrease in substructure at large $M/M_*$ may be manifested 
in our highest redshift data (Fig. \ref{vdfvirgrp1z}) where $M/M_* \sim 1000$.
We note however that our power law cosmology shows
that the substructure abundance is also strongly dependent on the
spectral slope.  The fact that low-mass haloes sample steeper
portions of the density power spectrum might cancel out any substructure
dependence on $M/M_*$, though the range of spectral slopes present
over our halo mass range is quite small.  Also, because low mass haloes
generally form earlier, one might expect them to have less substructure because
their subhaloes have been subject to tidal stripping and disruption for  
longer periods of time.

(5) {\it Is there a central subhalo ``anti-bias''?} The slope of the
subhalo number density is marginally consistent with that of the mass 
density for r$\simgt$0.3 ${\rm r_{vir}}$, although a subhalo distribution 
that is shallower than the dark matter profile is favoured.  More 
high-resolution simulations are needed to conclusively demonstrate a subhalo 
anti-bias at large radii.  
At smaller radii, subhaloes are deficient,
but artificial numerical effects cannot be excluded. It is not surprising
that more tidal stripping has occurred for subhaloes at small host radii,
as reflected in the mass-radius trend of Fig. \ref{momvsrpoints} and seen also
by \eg Gao \etal (2004a).
One would also expect that the removal of sufficiently large masses would
lead to a radial trend in ${\rm v_{cmax}}$, although such a radial trend
may be masked by the preferential destruction of low ${\rm v_{cmax}}$
haloes at small radii in our results. In any case, since subhaloes are
likely to have steep cuspy density profiles down to less than 1$\%$ ${\rm
r_{vir}}$ upon initial infall, as suggested by numerous studies of the
density profiles of virialized haloes (\eg Reed \etal 2005), then their
centres will be highly resistant to tidal disruption.  Thus, any
appearance of subhalo ``anti-bias'' in current generation simulations is
likely to be simply a manifestation of a radial trend in subhalo mass (or
${\rm v_{cmax}}$).  The imposition of an arbitrary minimum mass (or
minimum ${\rm v_{cmax}}$) as required by resolution constraints can give the
false impression of ``missing'' central subhaloes (see also Gao \etal 
2004b for a detailed discussion).  

(6) {\it Subhaloes have no strongly preferred orbits.} Orbital properties
show no strong trends with respect to radius, redshift, or ${\rm v_{cmax}}$.  
However, the subhaloes on the most highly eccentric orbits become less
abundant over time, likely due to disruption or heavy stripping by the 
central potential of
the host, but this only affects a small fraction of total subhaloes.  
Subhaloes with small apocentres are most strongly affected.
This likely has an artificial numerical cause wherein tidally affected 
subhaloes either lose enough material that they drop below resolution 
constraints, or are completely disrupted due to the effective
density ceiling imposed upon simulated subhaloes.
Our results also suggest that subhaloes have a
positive velocity bias at small radii and little or no velocity bias at
large radii.  It is not clear how the addition of baryons in the form of
gas and stars would affect orbital kinematics and distribution of the subhalo 
population.

(7) {\it Cosmological variance is too small to account for ``missing''
satellites.} The subhalo VDF has a halo to halo cosmological variance of a
factor of roughly two to four.  This means that the apparent problem of 
overpredicted local group satellites cannot be solved by invoking 
cosmological variance.

\section*{Acknowledgments} We would like to thank the referee for 
insightful suggestions which have improved this work.
We thank Lucio Mayer for assistance with one of
the runs.  We are grateful to Frank van den Bosch, Felix Stoehr, and Stelios
Kazantzidis for helpful suggestions upon reading a draft of this paper.
DR has been supported by the NASA Graduate Student Researchers
Program and by PPARC. FG is a David E. Brooks Research Fellow.  FG was
partially supported by NSF grant AST-0098557 at the University of
Washington. TRQ was partially supported by the National Science
Foundation. Simulations were performed on the Origin 2000 at NCSA and NASA
Ames, the IBM SP4 at the Arctic Region Supercomputing Center (ARSC) and at
CINECA (Bologna, Italy), the NASA Goddard HP/Compaq SC 45, and at the
Pittsburgh Supercomputing Center. We thank Chance Reschke for dedicated
support of our computing resources, much of which were graciously donated
by Intel.

{}

\label{lastpage}


\begin{thebibliography}{}

\bibitem{aubert} Aubert D., Pichon C., Colombi S., 2004, MNRAS, 352, 376

\bibitem{bard} Bardeen, J.M., Bond, J.R., Kaiser, N., Szalay, A.S., 1986,
ApJ, 305, 15.

\bibitem{barnes} Barnes J. E., Efstathiou G., 1987, ApJ, 319, 575

\bibitem{wmap} Bennett C. L. \etal, 2003, ApJS, 148, 1

\bibitem{bensona} Benson A., Lacey C., Baugh C., Cole S., Frenk C., 2002a, 
MNRAS, 333, 156B.

\bibitem{bensonb} Benson A., Frenk C., Lacey C., Baugh C., Cole S., 2002b,
MNRAS, 333, 177B.

\bibitem{kdtree} Bentley J. L., 1975, Communication of the ACM 18, 9

\bibitem{bullocksubreion1} Bullock J. S., Kravtsov A. V., Weinberg D. H., 2000,
ApJ, 539, 517.

\bibitem{bullockspin} Bullock J. S., Dekel A., Kolatt T., Kravtsov A., 
Klypin A., Porcianni C., Primack J., 2001, ApJ, 555, 240

\bibitem{chiba} Chiba M., 2002, ApJ, 565, 17

\bibitem{cole} Cole S., Lacey C., 1996, MNRAS, 281, 716

\bibitem{colin2000} Colin P., Klypin A., Kravtsov K., 2000, ApJ, 539, 561

\bibitem{colin} Colin P., Klypin A., Valenzuela O., Gottlober S.,
2004, ApJ, 612, 50

\bibitem{dalal} Dalal N., Kochanek C. S., 2002, ApJ, 572, 25

\bibitem{davis} Davis, M., Efstathiou, G., Frenk, C.S., White, S.D.M., 
1985, ApJ,
292, 381

\bibitem{delucia} De Lucia G., Kauffmann G., Springel V., White S. D. M., 
Lanzoni B., Stoehr F., Tormen G., Yoshida N., 2004, MNRAS, 348, 333

\bibitem{desai} Desai V., Dalcanton J. J., Mayer L., Reed D. S., Quinn T., 
Governato F., 2004, MNRAS, 351, 265

\bibitem{zurich2body} Diemand J., Moore B., Stadel J., Kazantzidis S.,
2004a, MNRAS, 348, 977

\bibitem{zsub} Diemand J., Moore B., Stadel J., 2004b, MNRAS, 352, 535

\bibitem{1996MNRAS...282..263}
Eke, V.R., Cole, S., Frenk, C.S., 1996, MNRAS, 282, 263

\bibitem{font} Font A. S., Navarro J. F., Stadel J., Quinn T., 2001, ApJ, 
563, L1

\bibitem{gaoa} Gao L., White S.D.M., Jenkins A., Stoehr F., Springel 
V., 2004a, MNRAS, 355, 819

\bibitem{gaob} Gao L., De Lucia G., White S.D.M., Jenkins A., 2004b, MNRAS, L1

\bibitem{gardner} Gardner J. P., 2001, ApJ, 557, 616

\bibitem{Ghignasub1} Ghigna S., Moore B., Governato F., Lake G., Quinn T.,
Stadel J., 1998, MNRAS, 300, 146

\bibitem{ghignasub2} Ghigna S., Moore B., Governato F., Lake G., Quinn
T., Stadel J., 2000, 544, 616

\bibitem{gillb} Gill P., Knebe A., Gibson B., Dopita M., 2004, MNRAS,
351, 410

\bibitem{gilla} Gill P., Knebe A., Gibson B., 2004a, MNRAS, 351, 399

\bibitem{gillb} Gill P., Knebe A., Gibson B., 2005, MNRAS, 356, 1327

\bibitem{gnedino} Gnedin O., 2003, ApJ, 582, 141

\bibitem{goto} Goto T., Yamauchi C., Fujita Y., Okamura S., Sekiguchi M., 
Smail I., Bernardi M., Gomez P., 2003, MNRAS, 346, 601

\bibitem{hayashisub} Hayashi, E., Navarro, J., Taylor, J., Stadel, J., 
Quinn, T, 2003, ApJ, 584, 541

\bibitem{katzrenormoverm} Katz N., White S., 1993, ApJ, 412, 478

\bibitem{zurichlgsats} Kazantzidis S., Mayer L., Mastropietro C., Diemand 
J., Stadel J., Moore B., 2004, ApJ, 608, 663

\bibitem{klyplg} Klypin A., Kravtsov A., Valenzuela O., and Prada F., ApJ, 
522, 82, 1999.

\bibitem{kravdark} Kravtsov A., Berlund A., Wechsler R., Klypin A., Gottlober
S., Allgood B., Primack J., 2004a, ApJ, 609, 35

\bibitem{krav} Kravtsov A., Gnedin O., Klypin A., 2004b, ApJ, 609, 482

\bibitem{lacey} Lacey C., Cole S., 1994, MNRAS, 271, 676

\bibitem{spinenv} Lemson G., Kauffmann G., 1999, MNRAS, 302, 111

\bibitem{maolens} Mao S., Schneider P, 1998, MNRAS, 295, 587

\bibitem{maolens2} Mao S., Jing Y., Ostriker J., Weller J., 2004, ApJ, 604, L5

\bibitem{mayer} Mayer L., Governato F., Colpi M., Moore B., Quinn T.,
Wadsley J., Stadel J., Lake G., 2001, ApJ, 547, L123

\bibitem{metcalflens} Metcalf, R., Madau, P., 2001, ApJ, 563, 9

\bibitem{mo} Mo H., Mao S., White S. D. M., 1998, MNRAS, 295, 319 

\bibitem{moore} Moore B., Katz N., Lake G., 1996, ApJ, 457, 455

\bibitem{m98} Moore B., Governato F., Quinn T., Stadel J., Lake G., 1998, 
AJ, 499, L5 

\bibitem{moorelg} Moore B., Ghigna S., Governato F., Lake G., Quinn T.,
Stadel J., Tozzi P., 1999, ApJ, 524, L19.

\bibitem{radsub} Nagai D., Kravtsov A., 2005, ApJ, 618, 557

\bibitem{peebles} Peebles J., 1969, ApJ, 155, 393

\bibitem{pie} Peirani S., Mohayaee R., Pacheco J., 2004, MNRAS, 348, 921

\bibitem{powerres} Power, C., Navarro, J. F., Jenkins, A., Frenk, C. S., White, S. D. 
M., Springel, V
., Stadal, J., \& Quinn, T.,
2003, MNRAS, 338, 14

\bibitem{ps} Press W.H., Schechter P., 1974, ApJ, 187, 425

\bibitem[Press \& Davis(1982)]{1982ApJ...259..449P} Press, W.~H.~\& Davis,
M.\ 1982, \apj, 259, 449

\bibitem{reed2003} Reed, D., Gardner, J., Quinn, T., Stadel, J., Fardal, M.,
Lake, G., \& Governato, F., 2003, MNRAS, 346, 565

\bibitem{reed2005} Reed, D., Governato, F., Verde, L., Gardner, J., 
Quinn, T., Merritt, D., Stadel, J., \& Lake, G., 2005, MNRAS, 357, 82

\bibitem{ryden} Ryden B. S., 1988, ApJ, 329, 589

\bibitem{1999MNRAS.308..119S} Sheth R.\ K.,
Tormen G., 1999, MNRAS, 308, 119

\bibitem{wmapn1} Spergel D., \etal, 2003, ApJS, 148, 175

\bibitem{som} Somerville R., 2002, ApJ, 572, L23

\bibitem{springelsub} Springel V., White S. D. M., Tormen G., Kauffmann G.,
2001, MNRAS, 328, 726

\bibitem{stadel} Stadel, J, 2001, PhDT.

\bibitem{stoehr} Stoehr F., White S. D. M., Tormen G., Springel V., 2002, MNRAS, 335, 84

\bibitem{stoehr2} Stoehr F., White S. D. M., Springel V., Tormen G., Yoshida N.,
2003, MNRAS, 345, 1313

\bibitem{syerwhiteunivpro} Syer D., White S. D. M., 1998, MNRAS, 293, 337

\bibitem{taffsats} Taffoni G., Mayer L., Colpi M., Governato F., 2003, 341, 434

\bibitem{taylor} Taylor J., Babul A., 2004, MNRAS, 348, 811

\bibitem{tormenstrip} Tormen G., Diaferio A., Syer D., 1998, MNRAS, 299, 728

\bibitem{vale} Vale A., Ostriker J., 2004, MNRAS, 353, 189

\bibitem{vandenbosch} van den Bosch F. C., Abel T., Croft R., Hernquist 
L., White S., 2002, ApJ, 576, 21

\bibitem{vandenboschcond} van den Bosch F., Yang X., Mo H., 2003, MNRAS, 340,
771

\bibitem{bandenboschbias} van den Bosch F., Norberg P., Mo H., Yang X., 2004, 
MNRAS, 352, 1302

\bibitem{verde} Verde L., Oh S. Peng, Jimenez R., 2002, MNRAS, 336, 541

\bibitem{vitvitska} Vitvitska M., Klypin A., Kravtsov A., Wechsler R., 
Primack J., Bullock J., 2002, ApJ, 581, 799

\bibitem{gas} Wadsley J., Stadel J., Quinn T., 2004, NewA, 9, 137

\bibitem{warren} Warren M., Quinn P., Salmon J., Zurek W, 1992 ApJ, 399, 
405

\bibitem{WhiteRees} White S. D. M., Rees M. J., 1978, MNRAS, 183, 341

\bibitem{morphrad} Whitmore B., Gilmore D., 1991, ApJ, 367, 64

\bibitem{wsub} Willman B., Governato F., Dalcanton J., Reed D., Quinn T., 
2004, MNRAS, 353, 639

\bibitem{yangcond} Yang X., Mo H., van den Bosch F., 2003, MNRAS, 339, 1057

\end{thebibliography}
\end{document}